\documentclass[aps,prb,reprint]{revtex4-1}
\usepackage{graphicx}
\usepackage{dcolumn}
\usepackage{bm}
\usepackage{hyperref}
\usepackage{amssymb}
\usepackage{amsmath}
\usepackage{epsf}
\usepackage[caption=false]{subfig}
\usepackage{epstopdf}
\usepackage{amsfonts}
\usepackage{color}

\newcommand{\mathbs}[1]{\mbox{\boldmath$#1$}}
\usepackage{pst-all}
\usepackage{graphicx}
\hbadness=\maxdimen
\vbadness=\maxdimen
\hfuzz=\maxdimen
\vfuzz=\maxdimen
\begin{document}
\title{Statistics of heat transport across capacitively coupled double quantum dot circuit}
\author{Hari Kumar Yadalam and Upendra Harbola}
\affiliation{Department of Inorganic and Physical Chemistry, Indian Institute of Science, Bangalore, 560012, India.}
\begin{abstract}
We study heat current and the full statistics of heat fluctuations in a capacitively-coupled double quantum dot system.  
This work is motivated by recent theoretical studies and experimental works on heat currents in quantum dot circuits. 
As expected intuitively, within the (static) mean-field approximation, the system at steady-state decouples into 
two single-dot equilibrium systems with renormalized dot energies, leading to zero average 
heat flux and fluctuations. This reveals that dynamic correlations induced between electrons on the dots is solely responsible 
for the heat transport between the two reservoirs. To study heat current fluctuations, we compute steady-state cumulant generating 
function for heat exchanged between reservoirs using two approaches : 
Lindblad quantum master equation approach, which is valid for arbitrary coulomb interaction strength but weak system-reservoir 
coupling strength, and the saddle point approximation for Schwinger-Keldysh coherent state  path integral, which is valid for arbitrary 
system-reservoir coupling strength but weak coulomb interaction strength. Using thus obtained generating functions, we verify 
steady-state fluctuation theorem for stochastic heat flux and study the average heat current and its fluctuations. We find that the heat current and 
its fluctuations change non-monotonically with the coulomb interaction strength ($U$) and system-reservoir coupling strength ($\Gamma$) 
and are suppressed for large values of $U$ and $\Gamma$.
\end{abstract}
\maketitle
\section{Introduction}
Studying transport processes in nano sized electronic quantum-dot junctions has been an active research area for last two decades 
\cite{Van2002,Joachim2005,Cuniberti2006,Tao2006,Cuevas2010,Baldea2016,Xiang2016}. The motivation being two fold : desire to design 
more efficient electronic devices and heat engines \cite{Sothmann2014}, and also as a platform for testing fundamental principles. From the technological 
perspective, useful devices have been proposed theoretically and few are tested experimentally. For example, nano diodes \cite{Aviram1974}, 
transistors \cite{Cardamone2006}, switches and other electronic elements relevant for device applications have been proposed. Understanding 
of charge and heat transport in nano systems are relevant for these applications. However, due to the small size, fluctuations of fluxes flowing through 
these systems are not negligible. These fluctuations are not arbitrary but follow universal relations called fluctuation theorems which 
generalize second law of thermodynamics to small scale 
\cite{Lebowitz1999,Jarzynski1997,Crooks1998,Crooks1999,Kurchan2000,Tasaki2000,Jarzynski2004,Monnai2005,Esposito2009,Campisi2011,Seifert2012}. 
These beautiful identities relate number 
of microscopic realizations of transport processes which produce certain amount of entropy to those which annihilate the same amount of entropy.  
Nano-electronic devices have served as useful platforms for testing these identities \cite{Utsumi2010,Kung2012,Pekola2015,Pekola2018}. These 
theorems are not only aesthetically appealing, but also are used to gain insights into transport processes. For example they have been 
used to characterize efficiency fluctuations \cite{Verley2014unlikely,Verley2014universal,Esposito2015efficiency} of nano heat engines, 
which is an important fundamental generalization of Carnot's analysis \cite{Callen2013} of macroscopic heat engines to micro scale. 
Thermoelectric engines, which are of current theoretical and experimental interest, constitutes one such class of nano heat 
engines \cite{Dubi2011,Sothmann2014,Perroni2016,Benenti2017,Cui2017} that convert heat to electrical work. 

Although heat flow plays a central role in determining the efficiencies of these engines, 
heat currents at nano electronic junctions are not as well explored as the charge currents. Recently there has been some 
interest in exploring the effects of various many-body interactions in thermoelectric heat engines. For example, effects of electron-phonon 
and electron-electron interactions on efficiencies of two-terminal and three-terminal thermoelectric engines are studied 
\cite{Azema2012,Zimbovskaya2014,Agarwalla2015,Perroni2015,Sierra2016,Thierschmann2016,Erdman2017,Dare2017,Friedman2017,Walldorf2017,Svilans2018,Daroca2018}. 
Furthermore, important experimental advancements in measuring heat currents in nano-electric junctions have been achieved recently 
\cite{Jezouin2013,Dutta2017,Cui2017}.

Motivated by these works, we study steady-state heat flux and fluctuations across capacitively (coulomb) 
coupled double quantum dot system. This system is known to act as a heat rectifier in some parameter regime \cite{Ruokola2011}. 
Capacitive coupling has been used to probe charge fluctuations in nano-junctions 
\cite{Cuetara2011,Golubev2011,Cuetara2015}, to understand coulomb drag effects, where electric charge flux in a circuit induces a charge flux in another 
capacitively coupled circuit \cite{Moldoveanu2009,Aita2013,Kaasbjerg2016,Keller2016,Narozhny2016,Zhou2019}. 
In  this work we are interested in the  study of heat-flux that is induced due to coulomb interactions between electrons in a capacitively coupled 
double-quantum dot system. To study heat flux and its fluctuations, we calculate cumulant generating function, defined using two point measurement scheme,  
using two different approaches valid in different parameter regimes (far above the Kondo-temperature \cite{Hewson1997}).  
Lindblad quantum master equation \cite{Breuer2002,Bagrets2003,Harbola2006,Harbola2007} valid at high temperatures, weak system reservoir coupling strength 
and arbitrary coulomb interaction strength, and Schwinger-Keldysh \cite{Rammer2007,Altland2010,Kamenev2011,Stefanucci2013} saddle-point method 
(random-phase approximation with mean-field dressed propagators) \cite{Hamann1969,Hamann1970,Morandi1974,Altland2010,Kamenev2011} valid for weak coulomb 
interaction strength and arbitrary system reservoir coupling strength. We verify steady-state heat fluctuation theorem and calculate heat flux and its 
fluctuations. Heat flux flowing through the same model system \cite{Wang2018} and fluctuations of heat flow in a variant of this model have been studied 
recently to  understand near-field radiative heat transfer within bare random-phase approximation \cite{Tang2018}. Here we present results that are valid beyond 
bare random-phase approximation (random phase approximation with mean-field dressed propagators).  We find that the steady-state scaled cumulant generating 
functions obtained using both the approximation schemes satisfy Gallavotti-Cohen symmetry and hence the steady-state fluctuation theorem for the 
heat fluctuations. Heat flux and its fluctuations are non-monotonic functions of coulomb interaction 
strength and decay exponentially for asymptotically large coulomb interaction strength. Similar non-monotonic behavior is seen with respect to 
system-reservoir coupling strength. The flux and its fluctuations are suppressed as a power law ($\Gamma_{}^{-4}$) for large coupling strength. 

In section II we introduce the model system. In section III we define moment generating function for heat fluctuations using two-point measurement scheme 
and calculate it using two approximation schemes and discuss heat flux and fluctuations in two subsections. We conclude in section IV.
\section{Model system}
\begin{figure}
\centering
\includegraphics[width=5.6cm,height=2.6cm]{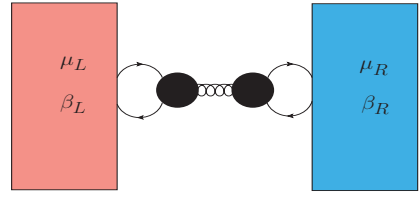}
\caption{(Color online) Schematic of the model considered.}
\label{model}
\end{figure}

Schematic of the model system considered in this work is shown in fig. (\ref{model}). It consists of two capacitively (Coulomb) coupled quantum dots (each 
having single orbital) individually coupled to two different fermionic reservoirs. The whole system is described by the following Hamiltonian, 
\begin{eqnarray}
\label{eq-1}
 \hat{H}&=&\underbrace{\sum_{\alpha=L,R}^{} \epsilon_{\alpha}^{} c_{\alpha}^{\dag} c_{\alpha}^{}}_{H_{S}^{}}^{} + U c_{L}^{\dag} c_{L}^{} c_{R}^{\dag} c_{R}^{} 
 + \sum_{\alpha=L,R}^{}\underbrace{\sum_{k}^{}\epsilon_{\alpha,k}^{} d_{\alpha k}^\dag d_{\alpha k}^{}}_{H_{\alpha}^{}}^{}\nonumber\\
 &&+ \mathop{\sum_k}_{\alpha=L,R}\big[g_{\alpha k}^{} d_{\alpha k}^\dag c_{\alpha}^{} + g_{\alpha k}^{*} c_{\alpha}^\dag d_{\alpha k}^{}\big].
\end{eqnarray}
Here  $c_{\alpha}^{\dag}$ ($c_{\alpha}^{}$) and $d_{\alpha k}^\dag$ ($d_{\alpha k}^{}$) stand for fermionic creation (annihilation) operator for creating 
(annihilating) an electron in the $\alpha^{th}$ ($\alpha_{}^{}=L,R$) quantum dot and in the state labeled by '$k$' in the $\alpha^{th}_{}$ fermionic reservoir 
respectively. The first term in Eq. (\ref{eq-1}) represents Hamiltonian of two isolated quantum dots each having a single orbital with energies 
$\epsilon_{\alpha}^{}$, the second term represents coulomb interaction between electrons on the two quantum dots, the third term is the Hamiltonian for the free 
electrons in the reservoirs, and the last term stands for hybridization between electrons on quantum dots and the reservoirs.
Throughout this work we assume wide-band approximation i.e., we assume that $g_{\alpha k}^{}$ is independent of $k$ and density of states of reservoirs are 
constant functions of energy. We note that the Hamiltonian given in Eq. (\ref{eq-1}) is a variant of the famous Anderson Hamiltonian 
\cite{Anderson1961,Anderson1978,Hewson1997}.

\section{Moment generating function}
When the quantum dots are brought together and are coupled to two reservoirs, energy and particles are exchanged. In this work, we are interested in 
calculating statistics of steady-state fluxes flowing through the double quantum dot system. In the long-time limit, only the heat flows between the 
left and the right reservoirs. The heat fluxes at the left and the right interface are balanced at the steady-state. The particle flux between system 
and reservoirs vanishes at steady-state. Physical reason for this is : (i) Since the coupling between the two quantum dots does not change number of 
particles on the dots (i.e., the particle exchange between the two dots is not allowed by microscopic dynamics), the net number of particles exchanged 
between the left(right) dot and the left(right) reservoir is constrained to $0$ and $1$, hence the particle flux and fluctuations are suppressed at 
the steady-state. (ii) Similarly energy cannot indefinitely accumulate on the system due to the boundedness of the systems energy spectrum, energy flux 
at the left and the right interfaces balance out at long-times.

Distribution function, $P[\Delta Q;T_{}^{}-T_{0}^{}]$, for heat ($\Delta Q$) flowing from the right reservoir to the left reservoir within a time 
$T_{}^{}-T_{0}^{}$ can be obtained using two-point measurement protocol \cite{Esposito2009,Campisi2011} for the observable corresponding to the operator 
$\frac{1}{2}[(H_{L}^{}-\mu_{L}^{}N_{L}^{})-(H_{R}^{}-\mu_{R}^{}N_{R}^{})]$ as,
\begin{eqnarray}
\label{eq-2}
 P[\Delta Q;T_{}^{}-T_{0}^{}]&=&\int_{-\infty}^{+\infty}\frac{d\chi_{}^{}}{2\pi}\mathcal{Z}[\chi;T_{}^{}-T_{0}^{}]e_{}^{i\chi_{}^{}\Delta Q},
\end{eqnarray}
where $\mathcal{Z}[\chi;T_{}^{}-T_{0}^{}]$ is the moment generating function for $\Delta Q$ and is given as,
\begin{eqnarray}
\label{eq-3}
 \mathcal{Z}[\chi;T_{}^{}-T_{0}^{}]&=&\mathbs{Tr}\big[\mathcal{U}_{\chi}^{}(T_{}^{},T_{0}^{})\rho(T_{0}^{})\mathcal{U}_{0}^{}(T_{0}^{},T_{}^{})\big],\nonumber\\
\end{eqnarray}
where $\mathcal{U}_{\chi}^{}(T_{1}^{},T_{2}^{})=e^{-\frac{i}{\hbar}(T_{1}^{}-T_{2}^{})H_{\chi}^{}}$ with 
\begin{eqnarray}
\label{eq-5}
&&H_{\chi}^{}=\sum_{\alpha=L,R}^{} \epsilon_{\alpha}^{} c_{\alpha}^{\dag} c_{\alpha}^{} + U c_{L}^{\dag} c_{L}^{} c_{R}^{\dag} c_{R}^{} 
+ \mathop{\sum_k}_{\alpha=L,R}\epsilon_{\alpha,k}^{} d_{\alpha k}^\dag d_{\alpha k}^{} \nonumber\\
&&+\mathop{\sum_k}_{\alpha=L,R}\big[g_{\alpha k}^{} e_{}^{-i(\epsilon_{\alpha k}^{}-\mu_{\alpha}^{})\chi_{\alpha}^{}} d_{\alpha k}^\dag c_{\alpha}^{} 
+ g_{\alpha k}^{*} e_{}^{i(\epsilon_{\alpha k}^{}-\mu_{\alpha}^{})\chi_{\alpha}^{}} c_{\alpha}^\dag d_{\alpha k}^{}\big],\nonumber\\
\end{eqnarray}
and $\chi_{L}^{}=-\chi_{R}^{}=\frac{\chi_{}^{}}{2}$. The trace in Eq. (\ref{eq-3}) is over the combined Fock space of the system and reservoirs and 
$\rho(T_{0}^{})$ is the density matrix of the whole system at initial time $T_{0}^{}$ taken here as 
\begin{eqnarray}
\label{eq-4}
\rho(T_{0}^{})&=&\frac{e_{}^{-\underset{\alpha=L,R,S}{\sum}\beta_{\alpha}\left[H_{\alpha}^{}-\mu_{\alpha}^{}N_{\alpha}^{}\right]}}
{Tr[e_{}^{-\underset{\alpha=L,R,S}{\sum}\beta_{\alpha}\left[H_{\alpha}^{}-\mu_{\alpha}^{}N_{\alpha}^{}\right]}]},
\end{eqnarray}
i.e., system and reservoirs initial states are non-interacting equilibrium states with different temperatures and chemical potentials.

Below we calculate $\mathcal{Z}[\chi;T_{}^{}-T_{0}^{}]$ approximately using two approaches, 
(i) Lindblad quantum master equation approach where coupling between system and reservoirs is assumed to be weak and, (ii) Schwinger-Keldysh path-integral 
approach where coulomb interaction strength is assumed to be weak but system reservoir coupling can be arbitrary.

\subsection{Lindblad quantum master equation approach}
$\mathcal{Z}[\chi;T_{}^{}-T_{0}^{}]$ defined in Eq. (\ref{eq-3}) can be expressed as,
\begin{eqnarray}
\label{eq-6}
 \mathcal{Z}[\chi;T_{}^{}-T_{0}^{}]&=&\mathbs{Tr}_{s}\big[\rho_{s}^{\chi}(T)\big],
\end{eqnarray}
where $\rho_{s}^{\chi}(T)=\mathbs{Tr}_{\text{Res}}\big[\mathcal{U}_{\chi}^{}(T_{}^{},T_{0}^{})\rho(T_{0}^{})\mathcal{U}_{\{0,0\}}^{}(T_{0}^{},T_{}^{})\big]$ 
is the counting field dependent reduced system density matrix at time $T$ obtained by tracing out the two reservoirs. Using the standard Born-Markov-Secular 
approximations (neglecting Lamb shifts), (counting field dependent) Lindblad quantum master equation can be derived for $\rho_{s}^{\chi}(T)$ 
\cite{Breuer2002,Bagrets2003,Harbola2006,Harbola2007}, which is given as,
\begin{widetext}
\begin{eqnarray}
\label{eq-7}
\frac{d}{dT}\rho_{s}^{\chi}(T)&=&-\frac{i}{\hbar}\sum_{\alpha=L,R}^{}\epsilon_{\alpha}^{}[ c_{\alpha}^\dag c_{\alpha}^{},\rho_{s}^{\chi}(T)]
-\frac{1}{2}\sum_{\alpha=L,R}^{}\Big[\Gamma_{\alpha}^{} f_{\alpha}^{}(\epsilon_{\alpha}+U c_{\bar{\alpha}}^\dag c_{\bar{\alpha}}^{}) 
\{c_{\alpha}^{} c_{\alpha}^\dag,\rho_{s}^{\chi}(T)\}+\Gamma_{\alpha}^{} [1-f_{\alpha}^{}(\epsilon_{\alpha}+U c_{\bar{\alpha}}^\dag c_{\bar{\alpha}}^{})] 
\{c_{\alpha}^\dag c_{\alpha}^{},\rho_{s}^{\chi}(T)\}\Big]\nonumber\\
&&+\sum_{\alpha=L,R}^{}\Big[\Gamma_{\alpha}^{} f_{\alpha}^{}(\epsilon_{\alpha}+U c_{\bar{\alpha}}^\dag c_{\bar{\alpha}}^{}) 
e_{}^{i(\epsilon_{\alpha}^{}-\mu_{\alpha}^{}+U c_{\bar{\alpha}}^\dag c_{\bar{\alpha}}^{})\chi_{\alpha}^{}} c_{\alpha}^\dag \rho_{s}^{\chi}(T) c_{\alpha}^{} 
+\Gamma_{\alpha}^{} [1-f_{\alpha}^{}(\epsilon_{\alpha}+U c_{\bar{\alpha}}^\dag c_{\bar{\alpha}}^{})] 
e_{}^{-i(\epsilon_{\alpha}^{}-\mu_{\alpha}^{}+U c_{\bar{\alpha}}^\dag c_{\bar{\alpha}}^{})\chi_{\alpha}^{}} 
c_{\alpha}^{}\rho_{s}^{\chi}(T) c_{\alpha}^\dag \Big],\nonumber\\
\end{eqnarray}
\end{widetext}
here $\alpha\neq\bar{\alpha}=L,R$; $\Gamma_{\alpha}^{}=\frac{2\pi}{\hbar}|g_{\alpha}^{}|^{2}_{} \rho_{\alpha}^{}$ and 
$f_{\alpha}^{}(\hat{o})=\left(e^{\beta_{\alpha}^{}(\hat{o}-\mu_{\alpha}^{})}_{}+1\right)_{}^{-1}$ with $\beta_{\alpha}^{}$ and $\mu_{\alpha}^{}$ being the 
temperature and chemical potential of the $\alpha_{}^{\text{th}}$ reservoir. 
It is important to note that if (static) mean-field approximation is made here, i.e., replacing $c_{\bar{\alpha}}^\dag c_{\bar{\alpha}}^{}$ by 
$\langle c_{\bar{\alpha}}^\dag c_{\bar{\alpha}}^{} \rangle_{S}^{}=\lim_{\left(T_{}^{}-T_{0}^{}\right)\to\infty}^{}\frac{\mathbs{Tr}_{s}
\big[c_{\bar{\alpha}}^\dag c_{\bar{\alpha}}^{}\rho_{s}^{\chi}(T)\big]}{\mathbs{Tr}_{s}\big[\rho_{s}^{\chi}(T)\big]}$, 
the right hand side of Eq. (\ref{eq-7}) can be separated into two terms which depend only on the dynamics of individual dots whose energies 
are renormalized by coupling to the other dot. This results in two decoupled quantum dots which equilibrate with their own reservoirs at long-time. 
Thus within this approximation, heat flux and fluctuations through the system vanish at steady-state. Hence mean-field approximation leads to no steady-state 
heat flux and fluctuations. And one needs to go beyond the mean-field approximation for having non-zero flux and fluctuations at steady state. 

By taking matrix elements in the occupation number basis of the two dots, $|N_{L}^{},N_{R}^{}\rangle$ (with $N_{\alpha}^{}= 0,1$), it can be seen that the 
populations ($\langle N_{L}^{},N_{R}^{}|\rho_{s}^{\chi}(T)|N_{L}^{},N_{R}^{}\rangle$) are decoupled from the coherences 
($\langle N_{L}^{},N_{R}^{}|\rho_{s}^{\chi}(T)|N_{L}',N_{R}'\rangle$), which die out exponentially fast with time.
Further, we restrict ourselves to a parameter regime : $\epsilon_{\alpha}^{}=\left(\mu_{\alpha}^{}-\frac{U}{2}\right)$, which simplifies the analysis. 
In this regime, we only need to solve the following $2 \times 2$ matrix equation, 
\begin{eqnarray}
\label{eq-8}
 \frac{d}{dT}|P_{s}^{\chi}(T)\rangle&=&\mathcal{L}_{}^{\chi}|P_{s}^{\chi}(T)\rangle,
\end{eqnarray}
where 
\begin{eqnarray}
|P_{s}^{\chi}(T)\rangle &=& \begin{bmatrix} \langle 1,0|\rho_{s}^{\chi}(T)|1,0\rangle + \langle 0,1|\rho_{s}^{\chi}(T)|0,1\rangle \\ 
\langle 0,0|\rho_{s}^{\chi}(T)|0,0\rangle + \langle 1,1|\rho_{s}^{\chi}(T)|1,1\rangle \end{bmatrix}
\end{eqnarray}
and the Liouvillian $\mathcal{L}_{}^{\chi}$ is given as,
\begin{eqnarray}
\label{eq-9}
\mathcal{L}_{}^{\chi}&=&\sum_{\alpha=L,R}^{} 
                          \begin{bmatrix} 
                           -\Gamma_{\alpha}^{} \bar{f}_{\alpha}^{} & \Gamma_{\alpha}^{} [1-\bar{f}_{\alpha}^{}]e_{}^{-i \frac{U}{2} \chi_{\alpha}^{}} \\
                            \Gamma_{\alpha}^{} \bar{f}_{\alpha}^{}e_{}^{i \frac{U}{2} \chi_{\alpha}^{}} & -\Gamma_{\alpha}^{} [1-\bar{f}_{\alpha}^{}]
                          \end{bmatrix},\nonumber\\
\end{eqnarray}
here $\bar{f}_{\alpha}^{}=\left(e_{}^{\beta_{\alpha}^{}U/2}+1\right)_{}^{-1}$. Note that the structure of the Liouvillian given in Eq. (\ref{eq-8}) is very similar to the 
case of charge transport through a resonant level system \cite{Goswami2015} when the two many-body states of the level are identified with the singly occupied 
and doubly the (un)occupied states of the double quantum dot system. 
Using solution of Eq. (\ref{eq-8}) in Eq. (\ref{eq-6}) (equivalent to 
$\mathcal{Z}[\chi;T_{}^{}-T_{0}^{}]=\langle \mathcal{I} | e_{}^{\mathcal{L}_{}^{\chi}(T_{}^{}-T_{0}^{})} | P_{s}^{0}(T_{0}^{}) \rangle$, 
with $\langle \mathcal{I} | = \begin{bmatrix} 1 & 1 \end{bmatrix}$), $\mathcal{Z}[\chi;T_{}^{}-T_{0}^{}]$ is obtained as,
\begin{widetext}
\begin{eqnarray}
\label{eq-10}
&&\mathcal{Z}[\chi;T_{}^{}-T_{0}^{}] 
         = e_{}^{-\frac{\left(\Gamma_{L}^{}+\Gamma_{R}^{}\right)}{2}(T_{}^{}-T_{0}^{})} 
         \Bigg\{\cosh\left[\Lambda_{}^{}[\chi_{}^{}]\left(T_{}^{}-T_{0}^{}\right)\right] + 
         \frac{\sinh\left[\Lambda_{}^{}[\chi_{}^{}]\left(T_{}^{}-T_{0}^{}\right)\right]}{\Lambda_{}^{}[\chi_{}^{}]} 
         \Bigg[\underset{\alpha=L,R}{\sum}\Gamma_{\alpha}^{}\Big(\frac{1}{2}\nonumber\\
         &&+\Big[f_{\alpha}^{}\left[f_{SL}^{}\left(1-f_{SR}^{}\right)+f_{SR}^{}\left(1-f_{SL}^{}\right)\right]
         \left(e_{}^{iU\chi_{\alpha}^{}/2}-1\right)+\left(1-f_{\alpha}^{}\right)\left[1-f_{SL}^{}\left(1-f_{SR}^{}\right)-f_{SR}^{}\left(1-f_{SL}^{}\right)
         \right]\left(e_{}^{-iU\chi_{\alpha}^{}/2}-1\right)\Big]\Big)\Bigg]\Bigg\},\nonumber\\
\end{eqnarray}
\end{widetext}
where $f_{S\alpha}^{}=\left(e_{}^{\beta_{S}^{}\left(\epsilon_{\alpha}^{}-\mu_{S}^{}\right)}+1\right)_{}^{-1}$ 
with $\beta_{S}^{}$ and $\mu_{S}^{}$ are temperature and chemical potential of the uncoupled quantum dots, and 
\begin{widetext}
\begin{eqnarray}
\label{eq-11}
 &&\Lambda_{}^{}[\chi_{}^{}]=\sqrt{\left(\frac{\Gamma_{L}^{}+\Gamma_{R}^{}}{2}\right)_{}^{2}+\Gamma_{L}^{}\Gamma_{R}^{}
 \left[\bar{f}_{L}^{}[1-\bar{f}_{R}^{}]\left(e_{}^{i U \chi_{}^{}/2}-1\right)+\bar{f}_{R}^{}[1-\bar{f}_{L}^{}]\left(e_{}^{- i U \chi_{}^{}/2}-1\right)\right]}.
 \end{eqnarray}
\end{widetext}
To arrive at explicit expression for $\mathcal{Z}[\chi;T_{}^{}-T_{0}^{}]$, we have used the initial condition i.e., 
\begin{eqnarray}
| P_{s}^{0}(T_{0}^{}) \rangle &=& \begin{bmatrix} f_{SL}^{}\left(1-f_{SR}^{}\right) + f_{SR}^{}\left(1-f_{SL}^{}\right) \\ 1 - f_{SL}^{}\left(1-f_{SR}^{}\right) 
- f_{SR}^{}\left(1-f_{SL}^{}\right)\end{bmatrix}_{}^{},
\end{eqnarray}
which is equivalent to $\rho_{s}^{\chi}(T_{0}^{})=\frac{e_{}^{-\beta_{S}\left[H_{S}^{}-\mu_{S}^{}N_{S}^{}\right]}}
{Tr[e_{}^{-\beta_{S}\left[H_{S}^{}-\mu_{S}^{}N_{S}^{}\right]}]}$.

In the long-time limit (i.e., $T_{}^{}-T_{0}^{} \to \infty$), the scaled cumulant generating function defined as,
\begin{eqnarray}
\label{eq-12}
 \mathcal{F}[\chi_{}^{}]&=&\lim_{(T_{}^{}-T_{0}^{}) \to \infty}^{} \frac{\ln\mathcal{Z}[\chi;T_{}^{}-T_{0}^{}]}{(T_{}^{}-T_{0}^{})}
\end{eqnarray}
is given by 

\begin{eqnarray}
\label{eq-13}
 \mathcal{F}[\chi_{}^{}] &=& -\frac{\left(\Gamma_{L}^{}+\Gamma_{R}^{}\right)}{2}+\Lambda_{}^{}[\chi_{}^{}].
\end{eqnarray}
This scaled cumulant generating function has the same form as that of for charge transport through a resonant level model \cite{Goswami2015}. 
This is due to the mapping between the two models as discussed earlier. 
It is straight forward to see that the cumulant generating function,  $\mathcal{F}[\chi_{}^{}]$, satisfies Gallavotti-Cohen symmetry : 
$\mathcal{F}[-\chi_{}^{}-i(\beta_{L}^{}-\beta_{R}^{})]=\mathcal{F}[\chi_{}^{}]$. 
This symmetry leads to the detailed steady-state fluctuation theorem for the distribution function for heat flow : 
$\lim_{(T_{}^{}-T_{0}^{}) \to \infty}^{} \frac{P[+\Delta Q;T_{}^{}-T_{0}^{}]}{P[-\Delta Q;T_{}^{}-T_{0}^{}]}=e_{}^{(\beta_{L}^{}-\beta_{R}^{})\Delta Q}$.

\begin{figure}
\centering
\includegraphics[width=7.2cm,height=4.5cm]{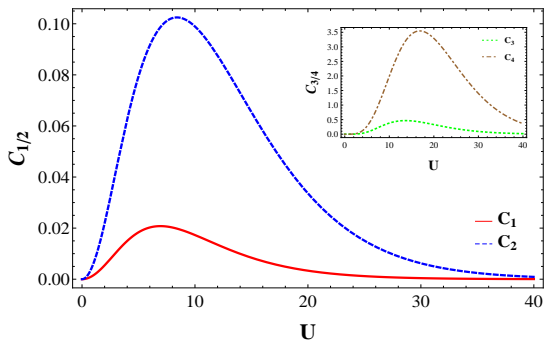}
\caption{(Color online) First and second cumulants of heat transfered from right to left reservoir as a function of coulomb interaction strength ($U$) for the 
parameters $\beta_{L}^{}=1.0$, $\beta_{R}^{}=0.5$, $\Gamma_{L}^{}=\Gamma_{R}^{}=0.1$. Inset shows third and fourth cumulants.}
\label{cumulants_lindblad}
\end{figure}

Further, using the above long-time limit scaled cumulant generating function, $\mathcal{F}[\chi_{}^{}]$, cumulants of heat flux can be obtained as 
$C_{n}^{}=i_{}^{n}\frac{d}{d\chi_{}^{n}}\mathcal{F}[\chi_{}^{}]$. Analytical expression for first four scaled cumulants, i.e, heat flux ($C_{1}^{}$), 
heat noise ($C_{2}^{}$), third cumulant ($C_{3}^{}$) and fourth cumulant ($C_{4}^{}$) are given as,
\begin{widetext}
\begin{eqnarray}
\label{eq-14}
C_{1}^{}&=& -\frac{U}{2}\frac{\Gamma_{L}^{}\Gamma_{R}^{}}{\Gamma_{L}^{}+\Gamma_{R}^{}}\left[\bar{f}_{L}^{}-\bar{f}_{R}^{}\right],\nonumber\\
C_{2}^{}&=& \left(\frac{U}{2}\right)_{}^{2}\frac{\Gamma_{L}^{}\Gamma_{R}^{}}{\Gamma_{L}^{}+\Gamma_{R}^{}}\left[\bar{f}_{L}^{}\left(1-\bar{f}_{R}^{}\right)
+\bar{f}_{R}^{}\left(1-\bar{f}_{L}^{}\right)\right]
-2\left(\frac{U}{2}\right)_{}^{2}\frac{\Gamma_{L}^{2}\Gamma_{R}^{2}}{\left(\Gamma_{L}^{}+\Gamma_{R}^{}\right)_{}^{3}}
\left[\bar{f}_{L}^{}-\bar{f}_{R}^{}\right]_{}^{2},\nonumber\\
C_{3}^{}&=& 6\left(\frac{U}{2}\right)_{}^{3}\frac{ \Gamma_{L}^{2} \Gamma_{R}^{2}}{(\Gamma_{L}^{}+\Gamma_{R}^{})_{}^{3}}(\bar{f}_{L}^{}-\bar{f}_{R}^{}) 
\left[\bar{f}_{L}^{}
   (1-\bar{f}_{R}^{})+\bar{f}_{R}^{}(1-\bar{f}_{L}^{})\right]-12\left(\frac{U}{2}\right)_{}^{3}
   \frac{\Gamma_{L}^{3} \Gamma_{R}^{3} }{(\Gamma_{L}^{}+\Gamma_{R}^{})_{}^{5}}(\bar{f}_{L}^{}-\bar{f}_{R}^{})_{}^{3}
   -\left(\frac{U}{2}\right)_{}^{3}\frac{\Gamma_{L}^{}\Gamma_{R}^{}}{(\Gamma_{L}^{}+\Gamma_{R}^{})}(\bar{f}_{L}^{}-\bar{f}_{R}^{}),\nonumber\\
C_{4}^{}&=&\left(\frac{U}{2}\right)_{}^{4}\frac{\Gamma_{L}^{}\Gamma_{R}^{}}{\Gamma_{L}^{}+\Gamma_{R}^{}}\left[\bar{f}_{L}^{}\left(1-\bar{f}_{R}^{}\right)
+\bar{f}_{R}^{}\left(1-\bar{f}_{L}^{}\right)\right]
-8\left(\frac{U}{2}\right)_{}^{4}\frac{\Gamma_{L}^{2}\Gamma_{R}^{2}}{\left(\Gamma_{L}^{}+\Gamma_{R}^{}\right)_{}^{3}}
\left[\bar{f}_{L}^{}-\bar{f}_{R}^{}\right]_{}^{2}
-120\left(\frac{U}{2}\right)_{}^{4}\frac{\Gamma_{L}^{4}\Gamma_{R}^{4}}{\left(\Gamma_{L}^{}+\Gamma_{R}^{}\right)_{}^{7}}
\left[\bar{f}_{L}^{}-\bar{f}_{R}^{}\right]_{}^{4}\nonumber\\
&&-6\left(\frac{U}{2}\right)_{}^{4}\frac{\Gamma_{L}^{2}\Gamma_{R}^{2}}{\left(\Gamma_{L}^{}+\Gamma_{R}^{}\right)_{}^{3}}
\left[\bar{f}_{L}^{}\left(1-\bar{f}_{R}^{}\right)+\bar{f}_{R}^{}\left(1-\bar{f}_{L}^{}\right)\right]_{}^{2}
+72\left(\frac{U}{2}\right)_{}^{4}\frac{\Gamma_{L}^{3}\Gamma_{R}^{3}}{\left(\Gamma_{L}^{}+\Gamma_{R}^{}\right)_{}^{5}}
\left[\bar{f}_{L}^{}-\bar{f}_{R}^{}\right]_{}^{2}\left[\bar{f}_{L}^{}\left(1-\bar{f}_{R}^{}\right)
+\bar{f}_{R}^{}\left(1-\bar{f}_{L}^{}\right)\right].\nonumber\\
\end{eqnarray}
\end{widetext}
Figure (\ref{cumulants_lindblad}) shows the four cumulants as a function of coulomb interaction strength. It is clear from this figure that heat flux 
and fluctuations are suppressed exponentially for large $U$. This is due to the exponential dependence of Fermi functions on $U$. Physically, the transition 
between the singly occupied states to doubly (un)occupied state become less probable as $U$ is increased \cite{Ruokola2011}. Further, we note that for 
intermediate values of $U$, fluctuations of heat are enhanced.

Since, $\mathcal{F}[\chi_{}^{}]$ is a periodic function of $\chi_{}^{}$ with period $\frac{4\pi}{U}$, this implies that 
$\underset{\left(T_{}^{}-T_{0}^{}\right)\to\infty}{\lim}P[\Delta Q;T_{}^{}-T_{0}^{}]$ has the Dirac comb structure : 
$\underset{\left(T_{}^{}-T_{0}^{}\right)\to\infty}{\lim}P[\Delta Q;T_{}^{}-T_{0}^{}]=\sum_{n=-\infty}^{+\infty}p[n;T_{}^{}-T_{0}^{}]
\delta[\Delta Q-\frac{n U}{2}]$ with $p[n;T_{}^{}-T_{0}^{}]=\int_{0}^{2\pi}\frac{d\chi}{2\pi}e_{}^{\mathcal{F}[\frac{2 \chi}{U}]
\left(T_{}^{}-T_{0}^{}\right)}e_{}^{i\chi n}$. 

\begin{figure}[tbh]
\centering
 \includegraphics[width=7.2cm,height=4.5cm]{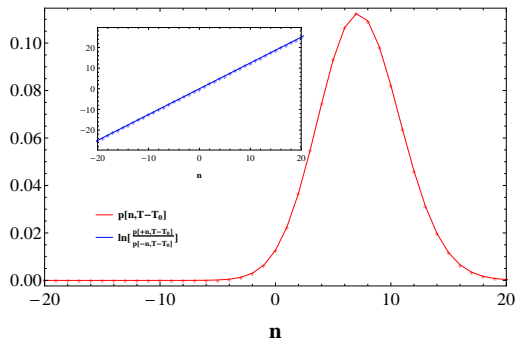}\\
 \caption{(Color online) Plot of $p[n,T_{}^{}-T_{0}^{}]$ vs '$n$' for $U=5.0$, 
 $T_{}^{}-T_{0}^{}=100.0\left(\frac{\Gamma_{L}^{}+\Gamma_{R}^{}}{2}\right)_{}^{-1}$ and all other parameters taken same as in fig. (\ref{cumulants_lindblad}). 
 Plot of $\ln\frac{p[n,T_{}^{}-T_{0}^{}]}{p[-n,T_{}^{}-T_{0}^{}]}$ vs '$n$' is shown in the inset.}
\label{probability_lindblad}
\end{figure}

 $p[n;T_{}^{}-T_{0}^{}]$ is computed numerically and is shown in fig. (\ref{probability_lindblad}) along with 
 $\ln\frac{p[n,T_{}^{}-T_{0}^{}]}{p[-n,T_{}^{}-T_{0}^{}]}$ in the inset, demonstrating the validity of steady-state Gallavotti-Cohen fluctuation theorem 
 for the stochastic heat flow.

In the next sub-section we present results obtained within saddle-point approximation for path-integral formulation of Schwinger-Keldysh technique.
\subsection{Schwinger-Keldysh path integral approach}
We compute the moment generating function ($\mathcal{Z}[\chi;T_{}^{}-T_{0}^{}]$ ) using path-integral on Schwinger-Keldysh contour. The results obtained are 
valid for arbitrary dot-reservoir coupling strength. However the effect of the coulomb interaction is incorporated approximately. 

\begin{figure}[tbh]
\centering
\includegraphics[width=7.0cm,height=3.8cm]{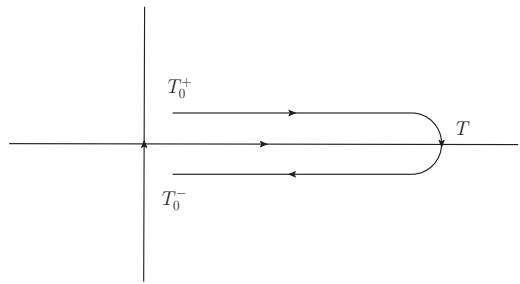}
\caption{Schematic of Schwinger-Keldysh contour.}
\label{contour}
\end{figure}

$\mathcal{Z}[\chi;T_{}^{}-T_{0}^{}]$, defined in Eq. (\ref{eq-3}), can be expressed as 
\begin{eqnarray}
\label{eq-15}
 \mathcal{Z}[\chi;T_{}^{}-T_{0}^{}]&=&\mathbs{Tr}\big[\mathcal{T}_{c}^{}e^{-\frac{i}{\hbar}\int_{c}d\tau H_{\chi_{}^{}(\tau)}^{}(\tau)}_{}
 \rho(T_{0}^{})\big],\nonumber\\
\end{eqnarray}
where $\mathcal{T}_{c}^{}e^{-\frac{i}{\hbar}\int_{c}d\tau H_{\chi_{}^{}(\tau)}^{}(\tau)}_{}$ is the evolution operator defined on the Schwinger-Keldysh contour 
\cite{Rammer2007,Haug2008,Kita2010,Kamenev2011,Stefanucci2013} shown in Fig. (\ref{contour}),  going from $T_{0}^{}$ to $T_{}^{}$ and back to $T_{0}^{}$. Here 
$H_{\chi_{}^{}(\tau)}^{}(\tau)=H_{\chi_{}^{}}^{}$ on forward contour and $H_{\chi_{}^{}(\tau)}^{}(\tau)=H_{\chi_{}^{}=0}^{}$ on the backward contour.
$ \mathcal{Z}[\chi;T_{}^{}-T_{0}^{}]$ can be expressed as a functional integral 
using Grassman field variables \cite{Negele1988,Rammer2007,Altland2010,Kamenev2011}, $\{\psi_{\alpha}^\dag(\tau),\psi_{\alpha}^{}(\tau)\}$ for the system and 
$\{\psi_{\alpha k}^\dag(\tau),\psi_{\alpha k}^{}(\tau)\}$ for the reservoirs. This gives,
\begin{eqnarray}
\label{eq-16}
  &&\mathcal{Z}[\chi;T_{}^{}-T_{0}^{}]=\frac{1}{\mathcal{N}}\int\mathcal{D}[\{\psi_{\alpha}^\dag(\tau),\psi_{\alpha}^{}(\tau)\}]\nonumber\\
  &&\int\mathcal{D}[\{\psi_{\alpha k}^\dag(\tau),\psi_{\alpha k}^{}(\tau)\}]e_{}^{\frac{i}{\hbar} S^{\chi_{}^{}}_{}[\{\psi_{\alpha}^\dag(\tau),
  \psi_{\alpha}^{}(\tau)\},\{\psi_{\alpha k}^\dag(\tau),\psi_{\alpha k}^{}(\tau)\}]},\nonumber\\
\end{eqnarray}
where $\mathcal{N}$ is the normalization constant (independent of $\chi$) such that $\mathcal{Z}[\chi;T_{}^{}-T_{0}^{}]|_{\chi=0}=1$. Here, we do not 
compute $\mathcal{N}$ explicitly, and modify  it at intermediate steps by absorbing all constants ('$\chi$' independent). Its value is 
determined finally by imposing $\mathcal{Z}[\chi;T_{}^{}-T_{0}^{}]|_{\chi=0}=1$.
$S^{\chi_{}^{}}_{}[\{\psi_{\alpha}^\dag(\tau),\psi_{\alpha}^{}(\tau)\},\{\psi_{\alpha k}^\dag(\tau),\psi_{\alpha k}^{}(\tau)\}]$ is the action of the 
whole system, given as,
\begin{widetext}
\begin{eqnarray}
\label{eq-17}
  &&S^{\chi_{}^{}}_{}[\{\psi_{\alpha}^\dag(\tau),\psi_{\alpha}^{}(\tau)\},\{\psi_{\alpha k}^\dag(\tau),\psi_{\alpha k}^{}(\tau)\}]\nonumber\\
  &&=\sum_{\alpha,\alpha'=L,R}\int_{c}^{}d\tau\int_{c}^{}d\tau' \Big[\psi_{\alpha}^\dag(\tau) [G_{sys}^{0}]^{-1}_{\alpha \alpha'}(\tau,\tau') 
  \psi_{\alpha'}^{}(\tau')+ \sum_{k,k'}^{} \psi_{\alpha k}^\dag(\tau) [G_{res}^{0}]^{-1}_{\alpha k \alpha' k'}(\tau,\tau') \psi_{\alpha' k'}^{}(\tau')
  \Big]\nonumber\\
  &&-\sum_{\alpha=L,R}^{}\sum_{k}^{} \int_{c}^{}d\tau \Big[g_{\alpha k}^{} e_{}^{-i(\epsilon_{\alpha k}^{}-\mu_{\alpha}^{})\chi_{\alpha}^{}(\tau)} 
  \psi_{\alpha k}^\dag(\tau) \psi_{\alpha}^{}(\tau) + g_{\alpha k}^{*} e_{}^{i(\epsilon_{\alpha k}^{}-\mu_{\alpha}^{})\chi_{\alpha}^{}(\tau)} 
  \psi_{\alpha}^\dag(\tau) \psi_{\alpha k}^{}(\tau)\big]-\int_{c}^{}d\tau U\psi_{L}^\dag(\tau)\psi_{L}^{}(\tau)\psi_{R}^\dag(\tau)\psi_{R}^{}(\tau),\nonumber\\
\end{eqnarray}
\end{widetext}
where  $\chi_{L}^{}(\tau)=-\chi_{R}^{}(\tau)=\frac{\chi_{}^{}}{2}$ on the forward contour and  $\chi_{L}^{}(\tau)=\chi_{R}^{}(\tau)=0$ on the backward contour. 
Further, $[G_{sys}^{0}]^{-1}_{\alpha \alpha'}(\tau,\tau')$ and $[G_{res}^{0}]^{-1}_{\alpha k \alpha' k'}(\tau,\tau')$ are matrix elements 
(with indices spanning state labels and contour times) of inverse of matrices with elements satisfying the following Schwinger-Dyson or Kadanoff-Baym equations 
on the Schwinger-Keldysh contour,
\begin{widetext}
 \begin{eqnarray}
 \label{eq-18}
  &&\sum_{\alpha_{1}^{}=L,R}\int_{c}^{}d\tau_{1}^{}\Big[\big(i\hbar\frac{\partial}{\partial \tau} -\epsilon_{\alpha}^{}\big)\delta_{\alpha \alpha_{1}^{}}^{}
  \delta_{}^{c}(\tau,\tau_{1}^{})\Big][G_{sys}^{0}]_{\alpha_{1}^{} \alpha'}(\tau_{1}^{},\tau')=\delta_{\alpha \alpha'}^{}\delta_{}^{c}(\tau,\tau')\nonumber\\
  &&\sum_{\alpha_{1}^{}=L,R}\int_{c}^{}d\tau_{1}^{}\Big[\big(-i\hbar\frac{\partial}{\partial \tau'} -\epsilon_{\alpha'}^{}\big)\delta_{\alpha_{1}^{} \alpha'}^{}
  \delta_{}^{c}(\tau_{1}^{},\tau')\Big][G_{sys}^{0}]_{\alpha \alpha_{1}^{}}(\tau,\tau_{1}^{})=\delta_{\alpha \alpha'}^{}\delta_{}^{c}(\tau,\tau')\nonumber\\
  &&\sum_{\alpha_{1}^{}=L,R}\sum_{k_{1}^{}}\int_{c}^{}d\tau_{1}^{}\Big[\big(i\hbar\frac{\partial}{\partial \tau} -\epsilon_{\alpha k}^{}\big)
  \delta_{\alpha \alpha_{1}^{}}^{}\delta_{k k_{1}^{}}^{}\delta_{}^{c}(\tau,\tau_{1}^{})\Big][G_{res}^{0}]_{\alpha_{1}^{} k_{1}^{} \alpha' k'}
  (\tau_{1}^{},\tau')=\delta_{\alpha \alpha'}^{}\delta_{k k'}^{}\delta_{}^{c}(\tau,\tau')\nonumber\\
   &&\sum_{\alpha_{1}^{}=L,R}\sum_{k_{1}^{}}\int_{c}^{}d\tau_{1}^{}\Big[\big(-i\hbar\frac{\partial}{\partial \tau'} -\epsilon_{\alpha' k'}^{}\big)
   \delta_{\alpha_{1}^{} \alpha'}^{}\delta_{k_{1}^{} k'}^{}\delta_{}^{c}(\tau_{1}^{},\tau')\Big][G_{res}^{0}]_{\alpha k \alpha_{1}^{} k_{1}^{}}
   (\tau,\tau_{1}^{})=\delta_{\alpha \alpha'}^{}\delta_{k k'}^{}\delta_{}^{c}(\tau,\tau')\nonumber\\
 \end{eqnarray}
\end{widetext}
with the following Kubo-Martin-Schwinger boundary conditions \cite{Martin1959,vanLeeuwen2013,Stefanucci2013} enforcing the information of the initial state of 
the system and reservoirs,
 \begin{eqnarray}
 \label{eq-19}
  &&[G_{sys}^{0}]_{\alpha \alpha'}(T_{0}^{-},\tau')=-e_{}^{\beta_{S}^{}\left(\epsilon_{\alpha}^{}-\mu_{S}^{}\right)}[G_{sys}^{0}]_{\alpha \alpha'}
  (T_{0}^{+},\tau')\nonumber\\
  &&[G_{sys}^{0}]_{\alpha \alpha'}(\tau,T_{0}^{-})=-e_{}^{-\beta_{S}^{}\left(\epsilon_{\alpha'}^{}-\mu_{S}^{}\right)}[G_{sys}^{0}]_{\alpha \alpha'}
  (\tau,T_{0}^{+})\nonumber\\
  &&[G_{res}^{0}]_{\alpha k \alpha' k'}(T_{0}^{-},\tau')=-e_{}^{\beta_{\alpha}^{}\left(\epsilon_{\alpha k}^{}-\mu_{\alpha}^{}\right)}
  [G_{res}^{0}]_{\alpha k \alpha' k'}(T_{0}^{+},\tau')\nonumber\\
  &&[G_{res}^{0}]_{\alpha k \alpha' k'}(\tau,T_{0}^{-})=-e_{}^{-\beta_{\alpha'}^{}\left(\epsilon_{\alpha' k'}^{}-\mu_{\alpha'}^{}\right)}
  [G_{res}^{0}]_{\alpha k \alpha' k'}(\tau,T_{0}^{+})\nonumber\\
 \end{eqnarray}
Equation (\ref{eq-19}) is one of the ways to take care of the initial state information in the Schwinger-Keldysh path integral formalism 
\cite{Utsumi2002,Altland2010,Kamenev2011,Secchi2018}. The solution of Eqs. (\ref{eq-18}) along with the boundary conditions Eqs. (\ref{eq-19}) are,
\begin{eqnarray}
\label{eq-20}
 &&[G_{sys}^{0}]^{}_{\alpha \alpha'}(\tau,\tau')=-\frac{i}{\hbar}e_{}^{-\frac{i}{\hbar}\epsilon_{\alpha}^{}\left(\tau-\tau'\right)}\nonumber\\
 &&\delta_{\alpha \alpha'}^{}
\left[\Theta(\tau,\tau')\left[1-f_{S}^{}(\epsilon_{\alpha}^{})\right]-\Theta(\tau',\tau)f_{S}^{}(\epsilon_{\alpha}^{})\right]\nonumber\\
&&[G_{res}^{0}]^{}_{\alpha k \alpha' k'}(\tau,\tau')=-\frac{i}{\hbar}e_{}^{-\frac{i}{\hbar}\epsilon_{\alpha k}^{}\left(\tau-\tau'\right)}\nonumber\\
&&
\delta_{\alpha \alpha'}^{}\delta_{k k'}^{}
\left[\Theta(\tau,\tau')\left[1-f_{\alpha}^{}(\epsilon_{\alpha k}^{})\right]-\Theta(\tau',\tau)f_{\alpha}^{}(\epsilon_{\alpha k}^{})\right]\nonumber\\
\end{eqnarray}
where $f_{X}^{}(x)=\left(e_{}^{\beta_{X}^{}(x-\mu_{X}^{})}+1\right)_{}^{-1}$ ($X=L,R,S$).
We integrate over the reservoir Grassman fields in Eq. (\ref{eq-16}) to get $\mathcal{Z}[\chi;T_{}^{}-T_{0}^{}]$ as a path integral only over system Grassman 
fields as,
\begin{eqnarray}
\label{eq-21}
  &&\mathcal{Z}[\chi;T_{}^{}-T_{0}^{}]=\nonumber\\
  &&\frac{1}{\mathcal{N}}\int\mathcal{D}[\{\psi_{\alpha}^\dag(\tau),\psi_{\alpha}^{}(\tau)\}] e_{}^{\frac{i}{\hbar} 
  S^{\chi_{}^{}}_{sys}[\{\psi_{\alpha}^\dag(\tau),\psi_{\alpha}^{}(\tau)\}]},
\end{eqnarray}
with 
\begin{eqnarray}
\label{eq-22}
  &&S^{\chi_{}^{}}_{sys}[\{\psi_{\alpha}^\dag(\tau),\psi_{\alpha}^{}(\tau)\}]=\nonumber\\
  &&\sum_{\alpha=L,R}\sum_{\alpha'=L,R}\int_{c}^{}d\tau\int_{c}^{}d\tau' \Big[\psi_{\alpha}^\dag(\tau) 
  [G_{}^{0}]^{-1}_{\alpha \alpha'}(\tau,\tau') \psi_{\alpha'}^{}(\tau')\Big]\nonumber\\
  &&-\int_{c}^{}d\tau U\psi_{L}^\dag(\tau)\psi_{L}^{}(\tau)\psi_{R}^\dag(\tau)\psi_{R}^{}(\tau),
\end{eqnarray}

where

\begin{eqnarray}
\label{eq-23}
 &&\sum_{\alpha_{1}^{}=L,R}\int_{c}^{}d\tau_{1}^{}\Big[[G_{sys}^{0}]^{-1}_{\alpha \alpha_{1}^{}}(\tau,\tau_{1}) -
 \Sigma_{\alpha \alpha_{1}^{}}^{c}(\tau,\tau_{1})\Big][G_{}^{0}]_{\alpha_{1}^{} \alpha'}(\tau_{1}^{},\tau')\nonumber\\
&&=\delta_{\alpha \alpha'}^{}\delta_{}^{c}(\tau,\tau'),
\end{eqnarray}
with the self-energy acquired by the system due to the coupling with the reservoirs given as,
\begin{eqnarray}
\label{eq-24}
\Sigma_{\alpha \alpha'}^{c}(\tau,\tau')=&&\delta_{\alpha \alpha'}^{}\sum_{k,k'}
g_{\alpha k}^{*}e_{}^{i\left(\epsilon_{\alpha k}^{}-\mu_{\alpha}^{}\right)\chi_{\alpha}^{}(\tau)}
g_{\alpha' k'}^{}e_{}^{-i\left(\epsilon_{\alpha' k'}^{}-\mu_{\alpha'}^{}\right)\chi_{\alpha'}^{}(\tau')} \nonumber\\
&&[G_{res}^{0}]^{}_{\alpha k \alpha k'}(\tau,\tau'), 
\end{eqnarray}

The path integral given in Eq. (\ref{eq-21}) over system Grassman fields cannot be evaluated exactly due to the presence of the quartic term 
(with coupling constant $U$). Hence, we proceed to evaluate it approximately. To that end, we decouple the above quartic term by introducing auxiliary 
real fields (aka Hubbard-Stratonovich decoupling) which can be interpreted as fluctuating external potentials 
\cite{Stratonovich1957,Hubbard1959,Kleinert1978,Negele1988,Altland2010,Kamenev2011}. This gives,
\begin{eqnarray}
\label{eq-25}
  &&\mathcal{Z}[\chi;T_{}^{}-T_{0}^{}]=\nonumber\\
  &&\frac{1}{\mathcal{N}}\int\mathcal{D}[\{\psi_{\alpha}^\dag(\tau),\psi_{\alpha}^{}(\tau)\}] 
  \int\mathcal{D}[\{\phi_{\alpha}^{}(\tau)\}]e_{}^{\frac{i}{\hbar} S^{\chi_{}^{}}_{}[\{\psi_{\alpha}^\dag(\tau),\psi_{\alpha}^{}(\tau)\},
  \{\phi_{\alpha}^{}(\tau)\}]},\nonumber\\
\end{eqnarray}
where 
\begin{eqnarray}
\label{eq-26}
&&S^{\chi_{}^{}}_{}[\{\psi_{\alpha}^\dag(\tau),\psi_{\alpha}^{}(\tau)\},\{\phi_{\alpha}^{}(\tau)\}]=\int_{c}^{} 
d\tau \frac{\phi_{L}^{}(\tau)\phi_{R}^{}(\tau)}{U} \nonumber\\
&&+\sum_{\alpha=L,R}\sum_{\alpha'=L,R}\int_{c}^{}d\tau\int_{c}^{}d\tau' \Big[\psi_{\alpha}^\dag(\tau) [G_{\phi_{}^{}}^{c}]^{-1}_{\alpha \alpha'}(\tau,\tau') 
\psi_{\alpha'}^{}(\tau')\Big],\nonumber\\
\end{eqnarray}

with $[G_{\phi_{}^{}}^{c}]^{-1}_{}(\tau,\tau')$ being the inverse of $[G_{\phi_{}^{}}^{c}]^{}_{}(\tau,\tau')$ which satisfy the following equation,
\begin{widetext}
\begin{eqnarray}
\label{eq-27}
 &&\sum_{\alpha_{1}^{}=L,R}\int_{c}^{}d\tau_{1}^{}\Big[[G_{sys}^{0}]^{-1}_{\alpha \alpha_{1}^{}}(\tau,\tau_{1})- \phi_{\alpha}^{}(\tau)
 \delta_{\alpha \alpha_{1}^{}}^{}\delta_{}^{c}(\tau,\tau_{1}^{}) -\Sigma_{\alpha \alpha_{1}^{}}^{c}(\tau,\tau_{1})\Big][G_{\phi}^{c}]_{\alpha_{1}^{} \alpha'}
 (\tau_{1}^{},\tau')=\delta_{\alpha \alpha'}^{}\delta_{}^{c}(\tau,\tau').
\end{eqnarray}
\end{widetext}

Since the Grassman path integral in Eq. (\ref{eq-25}) is quadratic in terms of system fields $\{\psi_{\alpha}^\dag(\tau),\psi_{\alpha}^{}(\tau)\}$, it can 
be performed exactly (here we have used the identity $\ln\det[\mathbb{A}]=\mathbs{Tr}[\ln \mathbb{A}]$) to get,

\begin{eqnarray}
\label{eq-28}
  \mathcal{Z}[\chi;T_{}^{}-T_{0}^{}]&=&\frac{1}{\mathcal{N}} \int\mathcal{D}[\{\phi_{\alpha}^{}(\tau)\}]e_{}^{\frac{i}{\hbar} 
  S^{\chi_{}^{}}_{}[\{\phi_{\alpha}^{}(\tau)\}]},\nonumber\\
\end{eqnarray}

with
\begin{eqnarray}
\label{eq-29}
 S^{\chi_{}^{}}_{}[\psi_{\alpha}^{}(\tau)\},\{\phi_{\alpha}^{}(\tau)\}]&=&\int_{c}^{}d\tau \frac{\phi_{L}^{}(\tau)\phi_{R}^{}(\tau)}{U}  
 -i\hbar\mathbs{Tr}_{}^{}\ln[[G_{\phi_{}^{}}^{c}]^{-1}],\nonumber\\
\end{eqnarray}
where $\mathbs{Tr}_{}^{}$ stands for trace over contour time and orbital indices. The above algebraic gymnastics doesn't solve the problem as the final 
path integral, Eq. (\ref{eq-28}), has an action, Eq. (\ref{eq-29}), which is highly non-linear, nevertheless it is a bosonic path integral, which 
can be approximately evaluated using saddle-point/stationary-phase method.
Within saddle-point approximation, action, $S^{\chi_{}^{}}_{}[\{\phi_{\alpha}^{}(\tau)\}]$ is functional Taylor expanded around the path 
$\{\phi_{\alpha}^{0}(\tau)\}$ which makes action stationary, i.e., 
\begin{eqnarray}
\label{eq-30}
&&\frac{\delta}{\delta \phi_{\alpha}^{}(\tau)}S^{\chi_{}^{}}_{}[\{\phi_{\alpha}^{}(\tau)\}]\Bigg|_{\{\phi_{\alpha}^{}(\tau)\}=\{\phi_{\alpha}^{0}(\tau)\}}^{}=0,
\end{eqnarray}
Further, the action is approximated by retaining terms in the functional Taylor expansions up to quadratic order, making the action functional, 
a quadratic form. This quadratic functional integral can be analytically evaluated to get a functional Fredholm determinant multiplied by the exponential of 
the action evaluated at the stationary path. With this hand-wavy description of saddle-point/stationary-phase approximation, we move ahead. 

The saddle point equations for the action given in Eq. (\ref{eq-29}) are obtained as,
\begin{eqnarray}
\label{eq-31}
 &&\phi_{L}^{0}(\tau)=-i\hbar U [G_{\phi_{}^{0}}^{c}]^{}_{R R}(\tau,\tau_{}^{+}) \nonumber\\
 &&\phi_{R}^{0}(\tau)=-i\hbar U [G_{\phi_{}^{0}}^{c}]^{}_{L L}(\tau,\tau_{}^{+})\nonumber\\
\end{eqnarray}
where an infinitesimal forward shift ($\tau_{}^{+}$) of the second argument compared to the first argument of 
$[G_{\phi_{x}^{0}}^{c}]^{}_{x x}(\tau,\tau_{}^{+})$ along the Schwinger-Keldysh contour can be deduced by consistently decoupling the fermionic quartic term 
using Hubbard-Stratonovich fields in discretized path integral. Otherwise there will be an ambiguity, as $[G_{\phi_{x}^{0}}^{c}]^{}_{x x}(\tau,\tau')$ is 
discontinuous at $\tau=\tau'$ with jump discontinuity of magnitude $\frac{i}{\hbar}$.  
Equations (\ref{eq-31}) for $\{\phi_{\alpha}^{0}(\tau)\}$ together with Eqs. (\ref{eq-27}) for $[G_{\phi_{}^{0}}^{c}]^{}_{\alpha \alpha'}(\tau,\tau')$, 
constitute self-consistent system of equations, which may possess more than one solution. When more than one stationary solution exist, then the functional 
integral is approximated by summing over the result obtained by Gaussian approximating the action around each of the stationary 
solutions. Expanding the action given in Eq. (\ref{eq-29}) around the stationary path and retaining only quadratic term, we get approximate expression for 
$\mathcal{Z}[\chi;T_{}^{}-T_{0}^{}]$ as (assuming that there is a unique stationary path),
\begin{eqnarray}
\label{eq-32}
  \mathcal{Z}[\chi;T_{}^{}-T_{0}^{}]&\approx& \frac{1}{\mathcal{N}} \int\mathcal{D}[\{\phi_{\alpha}^{}(\tau)\}]e_{}^{\frac{i}{\hbar} 
  S^{\chi_{}^{}}_{app}[\{\phi_{\alpha}^{}(\tau)\}]}\nonumber\\
\end{eqnarray}

with $S^{\chi_{}^{}}_{app}[\{\phi_{\alpha}^{}(\tau)\}]$ representing approximate action given as,
\begin{widetext}
\begin{eqnarray}
\label{eq-33}
 &&S^{\chi_{}^{}}_{app}[\psi_{\alpha}^{}(\tau)\},\{\phi_{\alpha}^{}(\tau)\}]=\int_{c}^{}d\tau \frac{\phi_{L}^{0}(\tau)\phi_{R}^{0}(\tau)}{U}  
 -i\hbar\mathbs{Tr}_{}^{}\ln[[G_{\phi_{}^{0}}^{c}]^{-1}]\nonumber\\
 &&+\frac{1}{2}\int_{c}^{}d\tau \int_{c}^{}d\tau' \begin{pmatrix}\phi_{L}^{}(\tau) - \phi_{L}^{0}(\tau) \\ \phi_{R}^{}(\tau) - 
 \phi_{R}^{0}(\tau) \end{pmatrix}_{}^{T}\left[ \begin{pmatrix}0 & \frac{\delta_{}^{c}(\tau,\tau')}{U} \\ \frac{\delta_{}^{c}(\tau,\tau')}{U} & 0\end{pmatrix} 
 + i\hbar \begin{pmatrix} P_{LL}^{0}(\tau,\tau') & 0 \\ 0 &  P_{RR}^{0}(\tau,\tau') \end{pmatrix}\right]\begin{pmatrix}\phi_{L}^{}(\tau') - \phi_{L}^{0}(\tau') 
 \\ \phi_{R}^{}(\tau') - \phi_{L}^{0}(\tau') \end{pmatrix}_{}^{},\nonumber\\
\end{eqnarray}
\end{widetext}
where $P_{\alpha \alpha}^{0}(\tau,\tau')=[G_{\phi_{}^{0}}^{c}]^{}_{\alpha \alpha}(\tau,\tau')[G_{\phi_{}^{0}}^{c}]^{}_{\alpha \alpha}(\tau',\tau)$ for 
$\alpha=L,R$ are the contour-ordered polarization propagators within random-phase approximation expressed in terms of mean-field system fermion propagators 
(solutions of Eqs. (\ref{eq-27}) and Eqs. (\ref{eq-31})). Note that the polarization dependent terms in Eq. (\ref{eq-33}) represent the leading order correction 
to the mean-field (saddle-point) contribution. After a change of variables (shift transformation $\{\phi_{\alpha}^{}(\tau)\}\to\{\phi_{\alpha}^{}(\tau)
+\phi_{\alpha}^{0}(\tau)\}$), followed by performing the final path integral over $\phi_{\alpha}^{}(\tau)$ and using the identity $\ln\det[\mathbb{A}]=
\mathbs{Tr}[\ln\mathbb{A}]$, we get,
\begin{widetext}
\begin{eqnarray}
\label{eq-34}
  &&\ln\mathcal{Z}[\chi;T_{}^{}-T_{0}^{}]\approx -\ln\mathcal{N} + \frac{i}{\hbar}\int_{c}^{}d\tau \frac{\phi_{L}^{0}(\tau)\phi_{R}^{0}(\tau)}{U} + 
  \mathbs{Tr}_{}^{}\ln[[G_{\phi_{}^{0}}^{c}]^{-1}]-\frac{1}{2}\mathbs{Tr}\ln\begin{pmatrix} i\hbar P_{LL}^{0}(\tau,\tau') & \frac{\delta_{}^{c}(\tau,\tau')}{U} 
  \\ \frac{\delta_{}^{c}(\tau,\tau')}{U} & i\hbar P_{RR}^{0}(\tau,\tau') \end{pmatrix}\nonumber\\
\end{eqnarray}
\end{widetext}
From now onwards we set $\hbar=1$. Further extracting $-\frac{1}{2}\ln\det \begin{pmatrix} 0 & U \delta_{}^{c}(\tau,\tau') \\ U \delta_{}^{c}(\tau,\tau') 
& 0 \end{pmatrix}$ from $\ln\mathcal{N}$ and combining with $-\frac{1}{2}\ln\det\begin{pmatrix} i P_{LL}^{0}(\tau,\tau') & \frac{\delta_{}^{c}(\tau,\tau')}{U} 
\\ \frac{\delta_{}^{c}(\tau,\tau')}{U} & i P_{RR}^{0}(\tau,\tau') \end{pmatrix}$ and using the identities, $\det\begin{pmatrix} \mathbb{I} & \mathbb{A} \\ 
\mathbb{B} & \mathbb{I} \end{pmatrix} = \det\left[\mathbb{I} -\mathbb{A} \mathbb{B} \right]$ and $\ln\det[\mathbb{A}]=\mathbs{Tr}[\ln \mathbb{A}]$, we get, 
\begin{widetext}
\begin{eqnarray}
\label{eq-35}
  &&\ln\mathcal{Z}[\chi;T_{}^{}-T_{0}^{}]\approx -\ln\mathcal{N} + i \int_{c}^{}d\tau \frac{\phi_{L}^{0}(\tau)\phi_{R}^{0}(\tau)}{U} + 
  \mathbs{Tr}_{}^{}\ln[[G_{\phi_{}^{0}}^{c}]^{-1}]-\frac{1}{2}\mathbs{Tr}_{}^{}\ln\left[\delta_{}^{c}(\tau,\tau') + U_{}^{2} \int_{c}^{}d\tau_{1}^{}
  P_{RR}^{0}(\tau,\tau_{1}^{})P_{LL}^{0}(\tau_{1}^{},\tau')\right]\nonumber\\
\end{eqnarray}
\end{widetext}
$\mathbs{Tr}$ in the above equation now stands only for trace over the contour time. The set of approximations made till now can be termed as mean-field dressed 
random-phase approximation based on the Feynman diagram representation of the final expression. 

Approximate expression for 
$\ln\mathcal{Z}[\chi;T_{}^{}-T_{0}^{}]$ given in Eq. (\ref{eq-35}) is valid for arbitrary measurement times $(T_{}^{}-T_{0}^{})$. But, in this work we are only 
interested in steady-state, hence we take $(T_{}^{}-T_{0}^{}) \to \infty$ and neglect information contained in the initial state of 
the system. For solving self consistent system of equations given in Eqs. (\ref{eq-27}) and Eqs. (\ref{eq-31}), we approximate, $\{\phi_{\alpha}^{0}(\tau)\}$ as 
independent of contour time ($\{\phi_{\alpha}^{0}(\tau)\}=\{\phi_{\alpha}^{}\}$) meaning, we assume that the stationary paths, $\{\phi_{\alpha}^{0}(\tau)\}$ 
as independent of time and are same on the forward and the backward branches of the contour. At this level, neglecting fluctuations of the Hubbard field, 
or stated equivalently, approximating the path integral within self-consistent Hartree-Fock/mean-field approximation leads to no heat flux and fluctuations 
at steady-state. This is because within this approximation only the second and the third terms (apart from normalization factor) which are independent of 
counting field are retained in Eq. (\ref{eq-35}). Hence fluctuations of Hubbard fields around their mean-field values are necessary to have finite 
heat flux and fluctuations. Within this 
approximation, equation for $[G_{\phi_{}^{}}^{c}]^{}_{\alpha \alpha'}(\tau,\tau')$, Eqs. (\ref{eq-27}), is solved in the frequency domain by first projecting 
it onto real times (which gives four Keldysh components for each $\alpha,\alpha'$) (notice that $[G_{\phi_{}^{}}^{c}]^{}_{\alpha \alpha'}(\tau,\tau') \propto 
\delta_{\alpha \alpha'}^{}$ is block diagonal in orbital space) and sending all temporal integrals from $-\infty$ to $+\infty$, followed by Fourier transforming 
to frequency domain \cite{Rammer2007,Haug2008}. The solution of Eq. (\ref{eq-27}) is then given as,
\begin{widetext}
\begin{eqnarray}
\label{eq-36}
 [G_{\phi_{}^{0}}^{}]^{}_{\alpha \alpha}(\omega) &=& \frac{1}{\left(\omega-\epsilon_{\alpha}^{}-\phi_{\alpha}^{}\right)_{}^{2}+
 \left(\frac{\Gamma_{\alpha}^{}}{2}\right)_{}^{2}}\begin{pmatrix} \left(\omega-\epsilon_{\alpha}^{}-\phi_{\alpha}^{}\right) - 
 i\frac{\Gamma_{\alpha}^{}}{2}\left(1-2 f_{\alpha}^{}(\omega)\right) & i\Gamma_{\alpha}^{}f_{\alpha}^{}(\omega)e_{}^{i\left(\omega-\mu_{\alpha}^{}\right)
 \chi_{\alpha}^{}} \\ -i\Gamma_{\alpha}^{}\left(1-f_{\alpha}^{}(\omega)\right)e_{}^{-i\left(\omega-\mu_{\alpha}^{}\right)\chi_{\alpha}^{}} & -
 \left(\omega-\epsilon_{\alpha}^{}-\phi_{\alpha}^{}\right) - i\frac{\Gamma_{\alpha}^{}}{2}\left(1-2 f_{\alpha}^{}(\omega)\right) \end{pmatrix},\nonumber\\
\end{eqnarray}
\end{widetext}
where $\Gamma_{\alpha}^{}=2\pi |g_{\alpha}^{}|_{}^{2} \rho_{\alpha}^{}$. With this, $[G_{\phi_{}^{0}}^{}]^{}_{\alpha \alpha}(\omega)$ is a function of 
constant stationary paths $\{\phi_{}^{0}\}$, which are determined self-consistently using Eq. (\ref{eq-31}), which in real-time read as,
\begin{eqnarray}
\label{eq-37}
  &&\phi_{L}^{0}=-i U [G_{\phi_{}^{0}}^{c}]^{++}_{R R}(t,t_{}^{+})=-i U [G_{\phi_{}^{0}}^{c}]^{--}_{R R}(t_{}^{+},t) \nonumber\\
 &&\phi_{R}^{0}=-i U [G_{\phi_{}^{0}}^{c}]^{++}_{L L}(t,t_{}^{+})=-i U [G_{\phi_{}^{0}}^{c}]^{--}_{L L}(t_{}^{+},t).\nonumber\\
\end{eqnarray}
Expressing these equations in frequency domain using Eq. (\ref{eq-36}), we get,
\begin{eqnarray}
\label{eq-38}
  &&\phi_{L}^{0}=U\int_{-\infty}^{+\infty}\frac{d\omega}{2\pi} \frac{\Gamma_{R}^{}f_{R}^{}(\omega)}{\left(\omega-\epsilon_{R}^{}-\phi_{R}^{0}\right)_{}^{2}
  +\left(\frac{\Gamma_{R}^{}}{2}\right)_{}^{2}}\nonumber\\
 &&\phi_{R}^{0}=U\int_{-\infty}^{+\infty}\frac{d\omega}{2\pi} \frac{\Gamma_{L}^{}f_{L}^{}(\omega)}{\left(\omega-\epsilon_{L}^{}-\phi_{L}^{0}\right)_{}^{2}
 +\left(\frac{\Gamma_{L}^{}}{2}\right)_{}^{2}}\nonumber\\
\end{eqnarray}
The $\omega$ integrals in the above equations can be analytically performed to get,
\begin{eqnarray}
\label{eq-39}
  &&\phi_{L}^{0}=U\left[\frac{1}{2}-\frac{1}{\pi}\textbf{Im}\Psi[\frac{1}{2}+\frac{\beta_{R}^{}\Gamma_{R}^{}}{4\pi}+i\frac{\beta_{R}^{}}{2\pi}
  \left(\epsilon_{R}^{}+\phi_{R}^{0}-\mu_{R}^{}\right)]\right]\nonumber\\
 &&\phi_{R}^{0}=U\left[\frac{1}{2}-\frac{1}{\pi}\textbf{Im}\Psi[\frac{1}{2}+\frac{\beta_{L}^{}\Gamma_{L}^{}}{4\pi}+i\frac{\beta_{L}^{}}{2\pi}
 \left(\epsilon_{L}^{}+\phi_{L}^{0}-\mu_{L}^{}\right)]\right]\nonumber\\
\end{eqnarray}
where $\textbf{Im}\Psi[z]$ is the imaginary part of digamma function evaluated at '$z$' \cite{Milne1972}. 
Eq. (\ref{eq-39}) are coupled non-linear self-consistent  equations for $\{\phi_{\alpha}^{0}\}$ which are difficult to solve analytically. 
However if we specialize to a special parameter regime $\epsilon_{\alpha}^{}=\mu_{\alpha}^{}-\frac{U}{2}$ and noting that $\textbf{Im}\Psi[z]=0$ for 
real $z$, it is clear that $\phi_{L}^{0}=\phi_{R}^{0}=\frac{U}{2}$ is always a stable solution for Eqs. (\ref{eq-39}), if  
\begin{eqnarray}
&&\sqrt{\frac{\beta_{L}^{}U}{2}}\sqrt{\frac{\beta_{R}^{}U}{2}} < \frac{\pi}{\sqrt{\Psi'[\frac{1}{2}+\frac{\beta_{L}^{}\Gamma_{L}^{}}
{4\pi}]}}\frac{\pi}{\sqrt{\Psi'[\frac{1}{2}+\frac{\beta_{R}^{}\Gamma_{R}^{}}{4\pi}]}}.\nonumber\\
\end{eqnarray}
Here onwards we confine ourselves to this regime. 

We simplify the expression for $\ln\mathcal{Z}[\chi;T_{}^{}-T_{0}^{}]$ given in Eq. (\ref{eq-35}) using the assumption that $\{\phi_{\alpha}^{0}(\tau)\}$ 
are independent of contour time, hence $\int_{c}^{}d\tau \frac{\phi_{L}^{0}(\tau)\phi_{R}^{0}(\tau)}{U}=0$ and absorbing $\mathbs{Tr}_{}^{}
\ln[[G_{\phi_{}^{0}}^{c}]^{-1}]$, which is independent of $\chi$ into $\ln\mathcal{N}$. Expanding the 
logarithmic term in Taylor series, projecting on to real times and sending intermediate time integrals to $-\infty$ to $+\infty$ and going over to the 
frequency domain, we get the long-time expression for scaled cumulant generating function as,
\begin{eqnarray}
\label{eq-40}
&&\mathcal{F}[\chi] \approx -\ln\mathcal{N} -\frac{1}{2}\int_{-\infty}^{+\infty}\frac{d\omega}{2\pi}\ln\det\left[\mathbb{I}_{2 \times 2}^{} 
+ U_{}^{2} P_{RR}^{0}(\omega)P_{LL}^{0}(\omega)\right].\nonumber\\
\end{eqnarray}
Here 
\begin{widetext}
\begin{eqnarray}
\label{eq-41}
 &&P_{\alpha \alpha}^{0}(\omega) = \int_{-\infty}^{+\infty}\frac{d\omega'}{2\pi}\begin{pmatrix} [G_{\phi_{}^{}}^{}]^{++}_{\alpha \alpha}
 (\omega+\omega')[G_{\phi_{}^{}}^{}]^{++}_{\alpha \alpha}(\omega') & [G_{\phi_{}^{}}^{}]^{+-}_{\alpha \alpha}(\omega+\omega')
 [G_{\phi_{}^{}}^{}]^{-+}_{\alpha \alpha}(\omega') \\ -[G_{\phi_{}^{}}^{}]^{-+}_{\alpha \alpha}(\omega+\omega')[G_{\phi_{}^{}}^{}]^{+-}_{\alpha \alpha}
 (\omega') & -[G_{\phi_{}^{}}^{}]^{--}_{\alpha \alpha}(\omega+\omega')[G_{\phi_{}^{}}^{}]^{--}_{\alpha \alpha}(\omega')\end{pmatrix}.
\end{eqnarray}
\end{widetext}
We notice that,
\begin{eqnarray}
\label{eq-42}
 P_{\alpha \alpha}^{0}(\omega) &=& \Lambda_{\alpha}^{}(\omega) \tilde{P}_{\alpha \alpha}^{0}(\omega)\Lambda_{\alpha}^{\dagger}(\omega),
\end{eqnarray}
where $\tilde{P}_{}^{}=P_{}^{}|_{\chi_{}^{}=0}^{}$ and $\Lambda_{\alpha}^{}(\omega)=e_{}^{i\frac{\chi_{\alpha}^{}}{2}\omega\sigma_{z}^{}}$ 
(where $\sigma_{z}^{}$ is the Pauli matrix). Further, $\tilde{P}_{\alpha \alpha}^{0}(\omega)$ can be expressed in terms of retarded, advanced and Keldysh 
projections of counting-field independent polarization propagators. After performing $\omega'$ integral in Eq. (\ref{eq-41}), we get,
\begin{eqnarray}
\label{eq-43}
 \tilde{P}_{\alpha \alpha}^{0}(\omega) &=& \mathcal{U}_{}^{T} \begin{pmatrix}[\tilde{P}_{\alpha \alpha}^{0}]^{R}_{}(\omega) & 
 [\tilde{P}_{\alpha \alpha}^{0}]^{K}_{}(\omega) \\ 0 & [\tilde{P}_{\alpha \alpha}^{0}]^{A}_{}(\omega) \end{pmatrix}\mathcal{U}_{}^{},
\end{eqnarray}
with $\mathcal{U}_{}^{}=\frac{1}{\sqrt{2}}\begin{pmatrix} 1 & -1 \\ 1 & 1 \end{pmatrix}$. Explicit expressions for counting field independent Keldysh 
rotated polarization propagators are given as : $[\tilde{P}_{\alpha \alpha}^{0}]^{A}_{}(\omega)=-
\left([\tilde{P}_{\alpha \alpha}^{0}]^{R}_{}(\omega)\right)_{}^{*}$ and $[\tilde{P}_{\alpha \alpha}^{0}]^{K}_{}(\omega)=
\left[1+2 n_{\alpha}^{}(\omega)\right]\left([\tilde{P}_{\alpha \alpha}^{0}]^{R}_{}(\omega)-[\tilde{P}_{\alpha \alpha}^{0}]^{A}_{}(\omega)\right)$ with 
$n_{\alpha}^{}(\omega)=\frac{1}{e_{}^{\beta_{\alpha}^{}\omega}-1}$ and 
\begin{widetext}
\begin{eqnarray}
\label{eq-44e}
 [\tilde{P}_{\alpha \alpha}^{0}]^{R}_{}(\omega)&=&\frac{i}{2\pi}\frac{\Gamma_{\alpha}^{}}{\omega+i\Gamma_{\alpha}^{}}
\frac{\Psi[\frac{1}{2}+\frac{\beta_{\alpha}^{}\Gamma_{\alpha}^{}}{4\pi}+i\frac{\beta_{\alpha}^{}}{2\pi}
 \left(\epsilon_{\alpha}^{}+\phi_{\alpha}^{}-\mu_{\alpha}^{}-\omega\right)]
 -\Psi[\frac{1}{2}+\frac{\beta_{\alpha}^{}\Gamma_{\alpha}^{}}{4\pi}+i\frac{\beta_{\alpha}^{}}{2\pi}
 \left(\epsilon_{\alpha}^{}+\phi_{\alpha}^{}-\mu_{\alpha}^{}\right)]}{\omega}\nonumber\\
 &+&\frac{i}{2\pi}\frac{\Gamma_{\alpha}^{}}{\omega+i\Gamma_{\alpha}^{}}
 \frac{\Psi[\frac{1}{2}+\frac{\beta_{\alpha}^{}\Gamma_{\alpha}^{}}{4\pi}-i\frac{\beta_{\alpha}^{}}{2\pi}
 \left(\epsilon_{\alpha}^{}+\phi_{\alpha}^{}-\mu_{\alpha}^{}+\omega\right)]
 -\Psi[\frac{1}{2}+\frac{\beta_{\alpha}^{}\Gamma_{\alpha}^{}}{4\pi}-i\frac{\beta_{\alpha}^{}}{2\pi}
 \left(\epsilon_{\alpha}^{}+\phi_{\alpha}^{}-\mu_{\alpha}^{}\right)]}{\omega}.
\end{eqnarray}
\end{widetext}
Using $\epsilon_{\alpha}^{}=\mu_{\alpha}^{}-\frac{U}{2}$ and $\phi_{L}^{0}=\phi_{R}^{0}=\frac{U}{2}$ in Eq. (\ref{eq-44e}) gives,
\begin{eqnarray}
\label{eq-44}
 [\tilde{P}_{\alpha \alpha}^{0}]^{R}_{}(\omega)&=&\frac{i}{\pi}\frac{\Gamma_{\alpha}^{}}{\left(\omega+i\Gamma_{\alpha}^{}\right)}
 \frac{\Psi[\frac{1}{2}+\frac{\beta_{\alpha}^{}\Gamma_{\alpha}^{}}{4\pi}-i\frac{\beta_{\alpha}^{}\omega}{2\pi}]-\Psi[\frac{1}{2}
 +\frac{\beta_{\alpha}^{}\Gamma_{\alpha}^{}}{4\pi}]}{\omega}.\nonumber\\
 \end{eqnarray}
 Before proceeding further, we note that $P_{\alpha \alpha}^{0}(\omega)$ given in Eqs. (\ref{eq-44e}) $\&$ (\ref{eq-44}) 
 is a meromorphic function with simple poles in the lower complex plane. Fourier transform (which can easily be obtained) 
 of Eq. (\ref{eq-44e}) display oscillations at characteristic frequency 
 $\left(\epsilon_{\alpha}^{}+\phi_{\alpha}^{}-\mu_{\alpha}^{}\right)$ and decay in time with rates depending linearly on 
 $\Gamma_{\alpha}^{}$ and $\beta_{\alpha}^{-1}$, whereas Fourier transform of Eq. (\ref{eq-44}) displays pure decay behavior, 
 as $\left(\epsilon_{\alpha}^{}+\phi_{\alpha}^{}-\mu_{\alpha}^{}\right)=0$.

Finally on using $P_{\alpha \alpha}^{0}(\omega)$ expressed above Eq. (\ref{eq-42}) in terms of $\Lambda_{\alpha}^{}(\omega)$ and Keldysh rotated quantities 
(Eq. (\ref{eq-43})) in Eq. (\ref{eq-40}) (and fixing $\ln\mathcal{N}$ by imposing normalization condition 
: $\ln\mathcal{Z}[\chi_{}^{}=0;T_{}^{}-T_{0}^{}]=0$), the final expression for long-time limit scaled cumulant generating function is obtained as,
\begin{widetext}
\begin{eqnarray}
\label{eq-45}
 \mathcal{F}[\chi]&=&-\frac{1}{2}\int_{-\infty}^{+\infty}\frac{d\omega}{2\pi}\ln\left[1-\mathbb{T}(\omega)\left(n_{L}^{}(\omega)
 \left[1+n_{R}^{}(\omega)\right]\left[e_{}^{i\chi_{}^{}\omega}-1\right]+n_{R}^{}(\omega)\left[1+n_{L}^{}(\omega)\right]
 \left[e_{}^{-i\chi_{}^{}\omega}-1\right]\right)\right]\nonumber\\
\end{eqnarray}
\end{widetext}
with the transmission function given by,
\begin{eqnarray}
\label{eq-46}
\mathbb{T}(\omega) &=& \frac{4 U_{}^{2}\mathbf{Re}\left([\tilde{P}_{L L}^{0}]^{R}_{}(\omega)\right)
\mathbf{Re}\left([\tilde{P}_{R R}^{0}]^{R}_{}(\omega)\right)}{\left|1+U_{}^{2}[\tilde{P}_{L L}^{0}]^{R}_{}(\omega)[\tilde{P}_{R R}^{0}]^{R}_{}(\omega)
\right|_{}^{2}}
\end{eqnarray}
where $[\tilde{P}_{\alpha \alpha}^{0}]^{R}_{}(\omega)$ are given in Eq. (\ref{eq-44}) and $n_{\alpha}^{}(\omega)$ is the bosonic distribution function. The 
algebraic form of the scaled cumulant generating function given in Eq. (\ref{eq-45}) is similar to that of heat transport across two-terminal bosonic 
harmonic junctions \cite{Saito2007,Agarwalla2012,Wang2014}. Unlike harmonic junctions, the transmission function given in Eq. (\ref{eq-46}) depends on 
temperatures of the reservoirs. We note that similar expression for scaled cumulant generating function (Eq. (\ref{eq-45})) for a variant of the model considered 
here \cite{Tang2018} and transmission function (Eq. (\ref{eq-46})) for the same model \cite{Wang2018} were obtained using bare random phase approximation 
recently. In contrast to these works, analytical expressions for polarization functions could be obtained here by invoking wide-band assumption. Using  
$P_{\alpha \alpha}^{0}(\omega)$ as defined in Eq. (\ref{eq-44e}) with $\phi_{L}^{0}=\phi_{R}^{0}=0$ in Eq. (\ref{eq-46}) gives transmission 
function obtained in Ref. (\cite{Wang2018}) for wide-band reservoirs case.

\begin{figure}
\centering
\includegraphics[width=7.2cm,height=4.5cm]{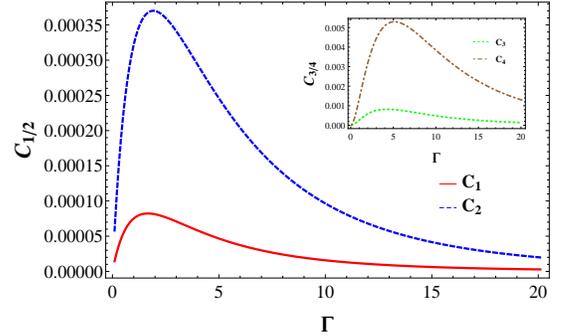}
\caption{(Color online) First and second cumulants of heat transfered from right to left reservoir as a function of system reservoir coupling strength 
($\Gamma_{L}^{}=\Gamma_{R}^{}=\Gamma_{}^{}$) for the parameters $\beta_{L}^{}=1.0$, $\beta_{R}^{}=0.5$, $U=0.1$. Inset shows third and fourth cumulants.}
\label{cumulants_keldysh}
\end{figure}

Expressions for first four long-time limit scaled cumulants ($C_{n}^{}=i_{}^{n}\frac{d_{}^{n}}{d\chi_{}^{n}} \mathcal{F}[\chi]$) are given as,
\begin{widetext}
\begin{eqnarray}
\label{eq-47}
C_{1}^{}&=& -\frac{1}{2}\int_{-\infty}^{+\infty} \frac{d\omega}{2\pi} \omega \mathbb{T}(\omega) \left[n_{L}^{}(\omega)-n_{R}^{}(\omega)\right] \nonumber\\
C_{2}^{}&=& \frac{1}{2}\int_{-\infty}^{+\infty} \frac{d\omega}{2\pi} \omega_{}^{2} \mathbb{T}(\omega)_{}^{2} \left[n_{L}^{}(\omega)-n_{R}^{}(\omega)\right]
_{}^{2}
+\frac{1}{2}\int_{-\infty}^{+\infty} \frac{d\omega}{2\pi} \omega_{}^{2} \mathbb{T}(\omega)_{}^{}\left[n_{L}^{}(\omega)\left(1+n_{R}^{}(\omega)\right)
+n_{R}^{}(\omega)\left(1+n_{L}^{}(\omega)\right)\right]_{}^{}\nonumber\\
C_{3}^{}&=& -\frac{1}{2}\int_{-\infty}^{+\infty} \frac{d\omega}{2\pi} \omega_{}^{3} \mathbb{T}(\omega) \left[n_{L}^{}(\omega)-n_{R}^{}(\omega)\right]
-\int_{-\infty}^{+\infty} \frac{d\omega}{2\pi} \omega_{}^{3} \mathbb{T}(\omega)_{}^{3} \left[n_{L}^{}(\omega)-n_{R}^{}(\omega)\right]_{}^{3}\nonumber\\
&&-\frac{3}{2}\int_{-\infty}^{+\infty} \frac{d\omega}{2\pi} \omega_{}^{3} \mathbb{T}(\omega)_{}^{2} \left[n_{L}^{}(\omega)-n_{R}^{}(\omega)\right]
\left[n_{L}^{}(\omega)\left(1+n_{R}^{}(\omega)\right)+n_{R}^{}(\omega)\left(1+n_{L}^{}(\omega)\right)\right]_{}^{}\nonumber\\
C_{4}^{}&=&3\int_{-\infty}^{+\infty} \frac{d\omega}{2\pi} \omega_{}^{4} \mathbb{T}(\omega)_{}^{4} \left[n_{L}^{}(\omega)-n_{R}^{}(\omega)\right]_{}^{4}
+6\int_{-\infty}^{+\infty} \frac{d\omega}{2\pi} \omega_{}^{4} \mathbb{T}(\omega)_{}^{3} \left[n_{L}^{}(\omega)-n_{R}^{}(\omega)\right]_{}^{2}
\left[n_{L}^{}(\omega)\left(1+n_{R}^{}(\omega)\right)+n_{R}^{}(\omega)\left(1+n_{L}^{}(\omega)\right)\right]_{}^{}\nonumber\\
&&+\frac{1}{2}\int_{-\infty}^{+\infty} \frac{d\omega}{2\pi} \omega_{}^{4} \mathbb{T}(\omega)_{}^{}
\left[n_{L}^{}(\omega)\left(1+n_{R}^{}(\omega)\right)+n_{R}^{}(\omega)\left(1+n_{L}^{}(\omega)\right)\right]_{}^{}
+\frac{7}{2}\int_{-\infty}^{+\infty} \frac{d\omega}{2\pi} \omega_{}^{4} \mathbb{T}(\omega)_{}^{2} \left[n_{L}^{}(\omega)-n_{R}^{}(\omega)\right]_{}^{2} 
\nonumber\\
&&+6\int_{-\infty}^{+\infty} \frac{d\omega}{2\pi} \omega_{}^{4} \mathbb{T}(\omega)_{}^{2}n_{L}^{}(\omega)n_{R}^{}(\omega)\left(1+n_{R}^{}(\omega)\right)
\left(1+n_{L}^{}(\omega)\right) \nonumber\\
\end{eqnarray}
\end{widetext}
Figure (\ref{cumulants_keldysh}) shows the plot long-time limit of first four scaled cumulants of heat flowing from right reservoir to the left reservoir as 
a function of system reservoir coupling strength ($\Gamma_{L}^{}=\Gamma_{R}^{}=\Gamma_{}^{}$). It can be seen that heat flux and fluctuations exhibit 
non-monotonic behavior as a function of system-reservoir coupling strength ($\Gamma_{L}^{}=\Gamma_{R}^{}=\Gamma_{}^{}$). This can be understood as follows 
: as noted previously electron density fluctuations on dots are solely responsible for heat flux, which are exponentially suppressed in time 
with rate depending linearly on $\Gamma_{}^{}$, hence the heat flux is suppressed for large $\Gamma_{}^{}$. At low temperatures 
($\beta_{L}^{-1}\approx0,\beta_{R}^{-1}\approx 0$), only low frequency behavior of transmission function is 
important, for small $\omega$, $\mathbb{T}(\omega)\approx\frac{16 U_{}^{2}}{\pi_{}^{2} \Gamma_{}^{4}}\omega_{}^{2}$ 
(here $\Gamma_{L}^{}=\Gamma_{R}^{}=\Gamma_{}^{}$). 
Hence at low temperatures and large system-reservoir coupling strength ($\Gamma_{}^{}$), heat current and fluctuations decay as a power law 
($\approx\Gamma_{}^{-4}$) with system-reservoir coupling strength. We note that similar non-monotonic behavior of particle flux through double-quantum dot 
system is observed recently \cite{Yadalam2018}. Further using this approximate expression for transmission function in the expression for heat flux 
($C_{1}^{}$ given in Eq. (\ref{eq-47})) we get the result \cite{Milne1972} 
$C_{1}^{}=\frac{16 \pi_{}^{2} U_{}^{2}}{15 \Gamma_{}^{4}}(\beta_{R}^{-4}-\beta_{L}^{-4})$ 
(Stefan-Boltzmann law \cite{Landau1975}) which is already noted in Ref. (\cite{Wang2018}).

We note that, $\mathcal{F}[-\chi_{}^{}-i\left(\beta_{L}-\beta_{R}^{}\right)]=\mathcal{F}[\chi_{}^{}]$, which is the steady-state Gallavotti-Cohen 
fluctuation symmetry, this symmetry leads to the standard steady-state fluctuation theorem for the stochastic heat flux flowing from the right 
reservoir to the left reservoir ($P[\Delta Q;T_{}^{}-T_{0}^{}]$) i.e., $\underset{(T_{}^{}-T_{0}^{}) \to \infty}{\lim_{}^{}} 
\frac{P[+\Delta Q;T_{}^{}-T_{0}^{}]}{P[-\Delta Q;T_{}^{}-T_{0}^{}]}=e_{}^{(\beta_{L}^{}-\beta_{R}^{})\Delta Q}$.

\begin{figure}[tbh]
\centering
 \includegraphics[width=7.2cm,height=4.5cm]{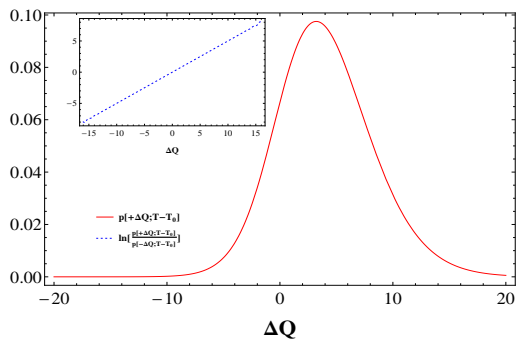}\\
 \caption{(Color online) Plot of $P[\Delta Q;T_{}^{}-T_{0}^{}]$ vs '$\Delta Q$' for $\Gamma_{L}^{}=\Gamma_{R}^{}=2.0$, 
 $T_{}^{}-T_{0}^{}=10_{}^{5}\left(\frac{\Gamma_{L}^{}+\Gamma_{R}^{}}{2}\right)_{}^{-1}$ and all other parameters taken same as in 
 fig. (\ref{cumulants_keldysh}). Plot of $\ln\frac{P[\Delta Q;T_{}^{}-T_{0}^{}]}{P[-\Delta Q;T_{}^{}-T_{0}^{}]}$ vs '$\Delta Q$' is shown in the inset.}
\label{probability_keldysh}
\end{figure}

Figure (\ref{probability_keldysh}) shows $P[\Delta Q;T_{}^{}-T_{0}^{}]$ and $P[-\Delta Q;T_{}^{}-T_{0}^{}]$ along with the inset plot of 
$\ln\frac{p[\Delta Q;T_{}^{}-T_{0}^{}]}{p[-\Delta Q;T_{}^{}-T_{0}^{}]}$ showing the validity of steady-state Gallavotti-Cohen fluctuation theorem.

\section{Conclusion}
In this work we studied heat flux and fluctuations of heat flowing across capacitively coupled double quantum dot circuit. We calculated moment generating 
function using two theoretical approaches valid in different parameter regimes. We found using Lindblad quantum master equation that heat flux and fluctuations 
exhibit non-monotonic behavior as a function of coulomb interaction strength and exponentially decay for strong coulomb interaction strength. Similarly 
using saddle point approximation scheme for Schwinger-Keldysh path integrals, 
heat flux and fluctuations are found to exhibit non-monotonic behavior as a function of system reservoir coupling strength and decay as inverse fourth power of 
system-reservoir coupling strength for large system-reservoir coupling strength. Further we have verified that the scaled cumulant generating function 
obtained using both the approximation schemes has Gallavotti-Cohen symmetry and hence the steady-state fluctuation theorem for the 
fluctuating heat flux is verified.
\section*{Acknowledgements}
H. Y. and U. H. acknowledge the financial support from the Indian Institute of Science  (India).
\section*{References}
\bibliography{Citation.bib}
\bibliographystyle{unsrt}

\end{document}